\documentclass[sigconf]{acmart}
\makeatletter
\renewcommand\@formatdoi[1]{\ignorespaces}
\makeatother
\renewcommand\footnotetextcopyrightpermission[1]{} 
\mathchardef\mhyphen="2D
\newcommand{\fontchange}[1]{\textsf{\small #1}}
\usepackage{multirow}
\usepackage{enumitem}

\AtBeginDocument{%
  }

\setcopyright{none}
\acmConference{C+J Conference 2022}{June 9 -- 11, 2022}{New York, NY, USA}
\begin{document}

\title[Characterizing Social Movement Narratives in Online Communities]{Characterizing Social Movement Narratives in Online Communities: The 2021 Cuban Protests on Reddit}

\author{Brian Keith Norambuena}
\email{briankeithn@vt.edu}
\affiliation{%
  \institution{Virginia Tech}
  \city{Blacksburg}
  \state{Virginia}
  \country{USA}
}
 \additionalaffiliation{%
   \institution{Universidad Católica del Norte}
   \department{Department of Computing \& Systems Engineering}
   \city{Antofagasta}
   \country{Chile}
   \postcode{1270709}
 }

\author{Tanushree Mitra}
\email{tmitra@uw.edu}
\affiliation{%
  \institution{University of Washington}
  \city{Seattle}
  \state{Washington}
  \country{USA}
}
\author{Chris North}
\email{north@vt.edu}
\affiliation{%
  \institution{Virginia Tech}
  \city{Blacksburg}
  \state{Virginia}
  \country{USA}
}
\renewcommand{\shortauthors}{Keith Norambuena et al.}

\begin{abstract}
Social movements are dominated by storytelling, as narratives play a key role in how communities involved in these movements shape their identities. Thus, recognizing the accepted narratives of different communities is central to understanding social movements. In this context, journalists face the challenge of making sense of these emerging narratives in social media when they seek to report social protests. Thus, they would benefit from support tools that allow them to identify and explore such narratives. In this work, we propose a narrative extraction algorithm from social media that incorporates the concept of community acceptance. Using our method, we study the 2021 Cuban protests and characterize five relevant communities. The extracted narratives differ in both structure and content across communities. Our work has implications in the study of social movements, intelligence analysis, computational journalism, and misinformation research. 
\end{abstract}


\maketitle

\section{Introduction}
Social movements are dominated by storytelling \cite{davis2002stories}. In fact, narratives play a key role in social change and social movements \cite{bell2010storytelling}, as people use them to learn and exercise agency, shape identity, and motivate actions \cite{dimond2013hollaback,ganz2001power}. Social movements are made possible by the construction of collective narratives within free spaces \cite{couto1993narrative}. In particular, online communities in social media provide free spaces for people to develop such narratives \cite{kow2016mediating}. Thus, studying how narratives evolve in these online communities is key to understanding social movements.

Journalists face unique challenges when reporting social movements \cite{mattoni2013journalism} and making sense of the emerging narratives in social media. In particular, the fluid nature and rapid evolution of social movements \cite{olessia2019explaining} makes it hard to keep track of its goals and motivations. In this context, identifying the driving narratives in the different communities involved in a social movement would aid journalists in their sensemaking process. The accepted narratives of a community provide a \textit{cognitive roadmap} by which community members are to live and define the collective identity of the group and their relations with other groups \cite{kaufman2009narratives, larsen2006comparative, kartchner2009strategic}, providing insight towards their motivations and goals. 

In this work, we propose a narrative extraction algorithm that incorporates the concept of \textit{community acceptance} to model and extract the accepted narratives from social media communities. Our proposed algorithm leverages the \textit{narrative maps} approach of Keith and Mitra \cite{keith2021narrative} to construct a complex narrative representation via optimization. The extracted narratives could provide journalists with an overview of the evolving narratives surrounding the social movement, giving them a window into these communities. In particular, we seek to answer the following research question: 
\begin{itemize}[leftmargin=*]
\itemsep0em 
\item \textbf{RQ}: How can we model community acceptance in social media and extract the accepted narratives of a social media community?
\end{itemize}

To test our method, we study the start of the \textit{2021 Cuban protests}---the biggest protests in decades in Cuba \cite{robles2021protests}---and the narratives surrounding this movement that emerged in online communities. Cuba has a unique sociopolitical, economic, and cultural landscape \cite{dye2018paquete}. Its long-standing communist regime faces unique challenges \cite{allahar2013bureaucratic}. The continued US embargo \cite{white2019ending} has undermined the Cuban economy \cite{kapcia2020short} and its health care system \cite{akbarpour2018impact}. The protests were triggered by food and medicine shortages \cite{pascal2021protests}, and the inefficient handling of the COVID-19 pandemic by the government \cite{faiola2021protests}.

Our approach should be able to capture these relevant events and storylines, allowing journalists to easily identify key elements of the narratives espoused by different communities in social media. In particular, we focus on a specific social media platform---Reddit---to explore how narratives change across communities. Specifically, we characterize the accepted narratives in a number of communities with a high level of activity related to the 2021 Cuban protests (e.g., \fontchange{r/cuba} and \fontchange{r/politics}). Thus, we make the following contributions:

\begin{enumerate}[leftmargin=*]
\itemsep0em 
    \item A narrative \textbf{extraction algorithm} that models \textit{community acceptance} in online communities, which could be used by journalists and other practitioners to explore social media narratives.
    \item An \textbf{analysis of the accepted narratives} espoused by online communities surrounding the 2021 Cuban protests to showcase the application of our method.
\end{enumerate}

We believe our approach has widespread implications, both for the study of social movement narratives online and narratives in general. At its core, our methods allow users to understand the accepted narratives of online communities. Furthermore, expert users, such as intelligence analysts, journalists, and misinformation researchers could use our methods to explore the narrative landscape of social movements emerging in online communities.

\section{Related Work}
Narratives are systems of stories with coherent themes \cite{halverson2011master}, where each story is represented as a sequence of events \cite{abbott2008cambridge}. Most narrative extraction methods use an event-based representation, where each event is associated with a single document in a data set \cite{shahaf2010connecting, keith2021narrative}. We follow this approach, as event-based representations have strong theoretical foundations in narratology \cite{abbott2008cambridge}. Narrative extraction approaches usually work by optimizing different criteria \cite{wang2015socially, shahaf2010connecting}. We use a narrative extraction algorithm based on the criteria of coherence maximization through linear programming, an extension of previous work in \cite{keith2021narrative}. However, unlike previous models, we integrate the concept of \textit{community acceptance} in our optimization model to select important events and storylines. 

Narratives are key to the formation of social movements \cite{bell2010storytelling,davis2002stories}. In this context, social media plays a key role in shaping the narrative \cite{kahne2016redesigning, kow2016mediating} and  mobilizing people \cite{leenders2012popular}. Thus, several works have sought to model these narratives. Computational methods to study social movement narratives in social media usually rely on keyword and hashtag analysis \cite{xiong2019hashtag}, topic modeling \cite{smoliarova2018detecting}, and user network analysis \cite{vicari2017twitter}. Other approaches include qualitative research, focusing on interviews and the experiences of people \cite{crivellaro2014pool,kou2017one}. However, none of these approaches consider the underlying structure of the narrative, which depends on temporal and causal relationships between events. In contrast, our approach extracts the narrative structure, allowing for a detailed model of the social movement narrative via narrative maps---an event-based directed acyclic graph narrative representation.

\section{Materials and Methods}
\textbf{Data Set Description.}
We used Pushshift \cite{baumgartner2020pushshift} to extract Reddit data. In particular, we extracted all submissions that contained the word ``Cuba'' from 07/07/2021 to 07/17/2021, covering a few days before the protests started on July 11th and the first week of the protests, allowing us to capture any useful signals on social media prior to the start of protests and the early narratives that emerged just after they started. We retrieved 4810 submissions. We focused on the top five communities according to their number of posts (see Table \ref{tab:communities}). We stopped at five because the sixth community (\fontchange{r/FreeKarma4U}) was a \textit{karma farming} group \cite{alizadeh2020content} unrelated to the protests that did not have a cohesive ideology or group identity and the number of posts rapidly declined in all subsequent communities. We manually analyzed all the submissions to ensure that they were relevant to the protests, finding that roughly 95\% out of 550 were relevant. 

\begin{table}[!htb]
\resizebox{\columnwidth}{!}{%
\begin{tabular}{@{}lp{8cm}ll@{}}
\toprule
\textbf{Community} & \textbf{Description}                                            & \textbf{Posts} & \textbf{Subscribers} \\ \midrule
\fontchange{r/cuba}             & The biggest Cuban community in Reddit.                          & 168 & 36k
         \\
\fontchange{r/Conservative}     & US-centric right-leaning community focused on political issues. & 126 & 833k
         \\
\fontchange{r/GenZedong}        & Pro-communist community focused on humor and world news.        & 99  & 33k           \\
\fontchange{r/worldnews}        & Community focused on world news (excluding US news).            & 79  & 26.6M           \\
\fontchange{r/politics}         & US-centric left-leaning community focused on political issues.  & 78  & 7.6M
           \\ \bottomrule
\end{tabular}%
}
\caption{Summary of the analyzed Reddit communities, including the number of subscribers and submissions retrieved.}
\label{tab:communities}
\vspace{-20pt}
\end{table}

\textbf{Narrative Extraction Method.}
We use narrative maps \cite{keith2021narrative} as our representation (see Figure \ref{fig:cuba}). Narrative maps provide a generic foundation to encode different types of narratives extracted from data, requiring only the existence of a total ordering (e.g., timestamps) and text representation of the event (e.g., a news headline or post title). Based on a \textit{route map} metaphor, they consist of a directed acyclic graph of event nodes, where each path represents a route connecting events. These nodes are called \textit{landmarks}, based on the idea of route maps having a series of landmarks to highlight important elements. Certain events are highlighted---called \textit{representative landmarks}---at the point of highest width in the graph to capture all parallel storylines. Finally, by finding the path of maximum coherence we find the \textit{main route} that contains the core events of the narrative. Narrative maps are extracted by optimizing coherence relations between events---based on event similarity and topical consistency---subject to coverage constraints. 

In this work, we propose a new optimization model that accounts for \textit{community acceptance} in addition to coherence and story coverage. We include new parameters to account for the \textbf{submission score} (i.e., the difference between upvotes and downvotes) and \textbf{upvote ratio} as a way to measure \textbf{community acceptance} of a submission. In particular, posts that have a high score and upvote ratio in a subreddit are considered accepted by that community \cite{buz2021should, li2020race} and should appear in the narrative. We model this with a constraint that requires the average score percentile to be above a certain threshold. Note that this does not necessarily discard all the low score submissions, as their presence might be necessary to construct a coherent narrative. Furthermore, we incorporate \textbf{community acceptance} into the optimization objective itself, extracting maps by maximizing both coherence and acceptance. 

\begin{figure}[!htb]
\centering
\includegraphics[width=0.75\columnwidth]{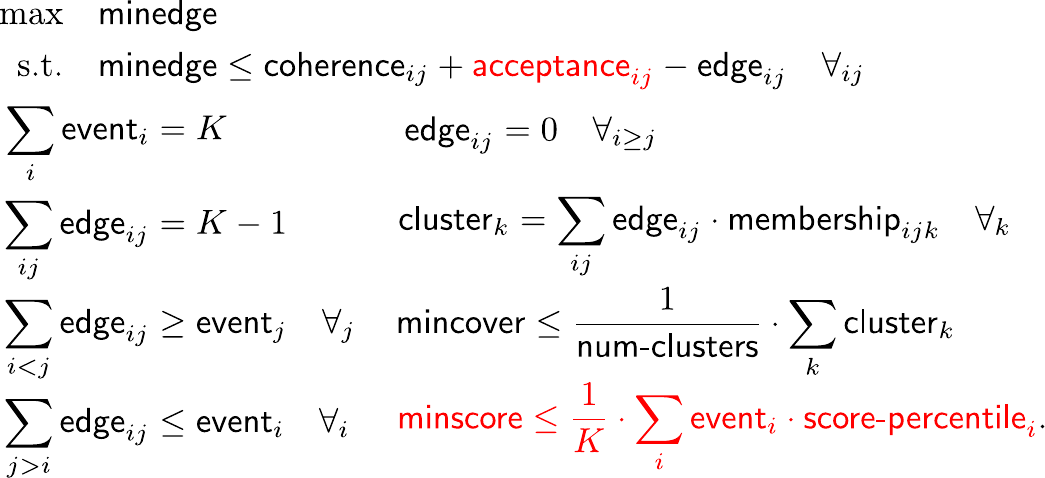}
\caption{Linear program formulation. Parts highlighted in red are our contributions to the model.}
\label{fig:lp}
\vspace{-13pt}
\end{figure}

We show the optimization model in Figure \ref{fig:lp}. Our goal is to maximize the minimum edge strength ($\mathsf{minedge}$), which depends on $\mathsf{coherence}$---a measure of how much sense it makes to join two events together, based on event similarity and topical similarity---and $\mathsf{acceptance}$---a measure of community acceptance that depends on the upvote ratios and submission score in the community. Table \ref{tab:lp_model} in the appendix presents the details of our model.

For the parameters, we set $\mathsf{mincover} = 0.5$, ensuring that we cover at least half of the stories without excessively restricting the narrative. We set $K = 8$ to get main stories with around 8 events, based on \cite{keith2021narrative, shahaf2010connecting}. We set $\mathsf{minscore} = 0.85$, ensuring that the average community acceptance is at least 85\%---intuitively, this means that roughly 85\% of the community regards the selected submissions positively. Lower thresholds allowed too many low-acceptance events to be on the map. We encode submissions with Multilingual USE embeddings \cite{yang2019multilingual}, as some communities use English and Spanish. The rest of the implementation follows \cite{keith2021narrative}. We note that all parameters can be easily modified, allowing journalists and other users to explore their influence on the narrative. 

\section{Results}
Now we present the accepted narratives extracted from each community. Due to space limits, we offer details of only one community (\fontchange{r/cuba}) while briefly mentioning the rest. We also analyze how the narratives espoused by each community differ from each other and manually categorize the relevant storylines and events.

\begin{figure*}[!htb]
    \centering
    \includegraphics[width=0.63\textwidth]{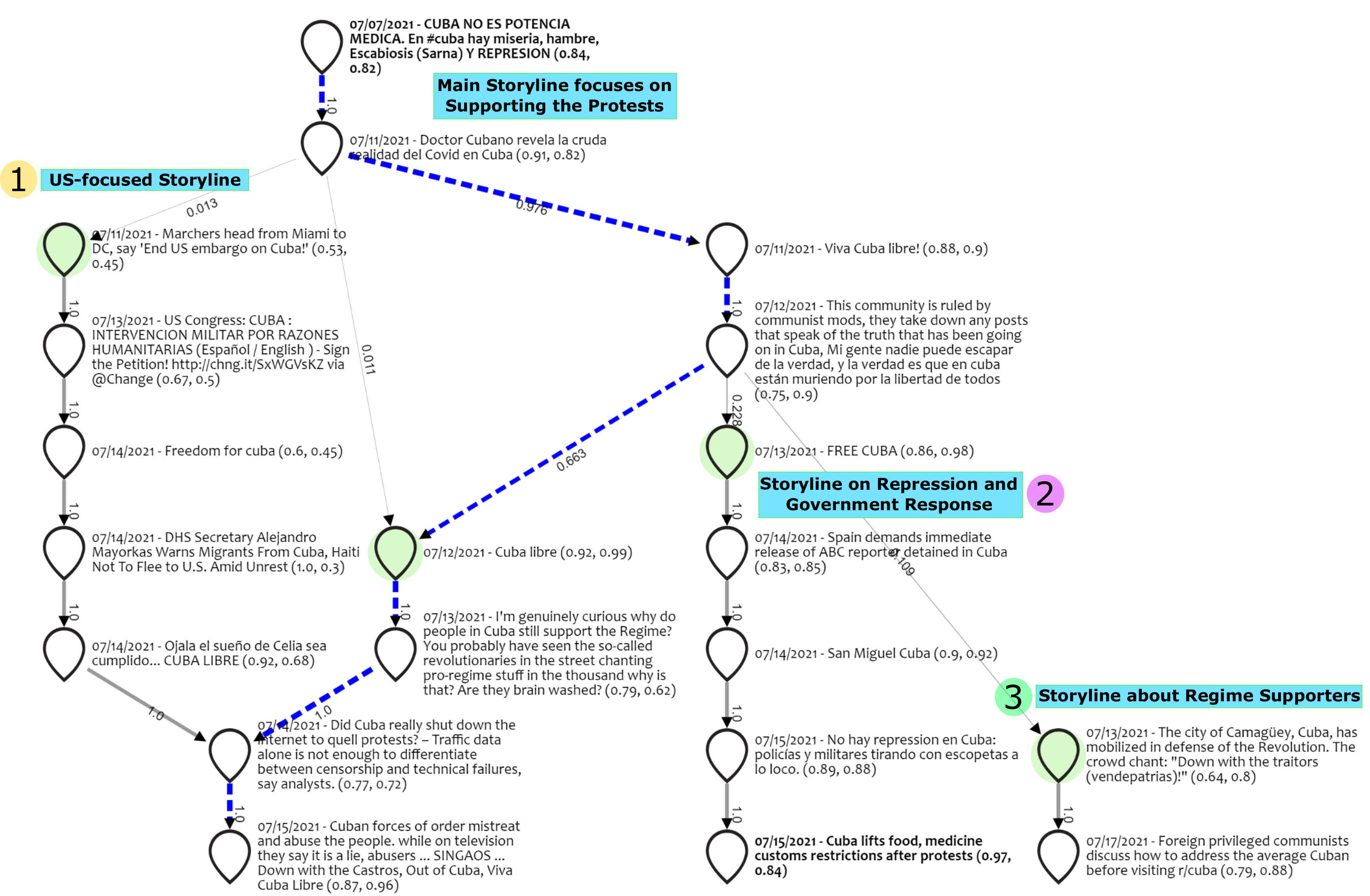}
    \caption{Narrative map for \fontchange{r/cuba}. The numbers beside each submission represent the upvote ratio and score percentile. The main storyline is represented by the blue dashed lines. Highlighted nodes correspond to representative landmarks \protect\resizebox{0.30cm}{!}{\protect\includegraphics{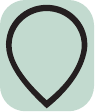}}. Edge weights represent coherence. Manual annotations describing specific storylines have been added with markers \protect\resizebox{0.30cm}{!}{\protect\includegraphics{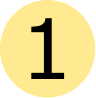}}, \protect\resizebox{0.30cm}{!}{\protect\includegraphics{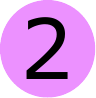}}, and \protect\resizebox{0.30cm}{!}{\protect\includegraphics{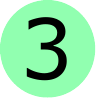}}.}
    \label{fig:cuba}
    \vspace{-10pt}
\end{figure*}

\textbf{\fontchange{r/cuba}:}
Figure \ref{fig:cuba} shows the narrative map extracted from \fontchange{r/cuba}. The main storyline starts with a submission criticizing the standing of Cuba as a leader in health care. Moreover, it highlights economical issues and political repression. Next, a submission discusses the state of the pandemic in Cuba on the same day that the protests start. Overall, the main storyline gives the impression that the protests started due to health care issues, alongside other catalysts. 

Afterward, we see a simple submission with the rallying call ``Viva Cuba libre!'' (translated as ``long live free Cuba'') with an image about tyranny in its internal contents. We note that this submission was locked by the moderators of the community. The next day there is a post criticizing the moderators for taking up anything about the protests. Then, we see a rallying call with a video showing the protests. After this, there are two submissions seeking information about the protests, asking about counter-protests and whether the internet has been shut down. Finally, the story ends with reports of abuse by Cuban police and a call against the regime.

Second, we analyze the representative landmarks highlighted in green and their storylines. Storyline \resizebox{0.30cm}{!}{\includegraphics{figures/1.pdf}} and its representative landmark are focused on the US. This story discusses the US embargo, a potential intervention, and a migratory crisis. The second representative landmark is a rallying call in the main storyline, which makes sense as a key narrative event that highlights the protests. The third representative landmark is another rallying call with a photo of protesters carrying a Cuban flag. The rest of the storyline \resizebox{0.30cm}{!}{\includegraphics{figures/2.pdf}} focuses mostly on repression in the protests, with Spain demanding the release of a reporter, and the police and military shooting protesters. However, it ends on a better note with changes in customs restrictions by the Cuban government due to the protests. The final representative landmark and side storyline \resizebox{0.30cm}{!}{\includegraphics{figures/3.pdf}} focus on the counter-protests and support for the regime, with people mobilizing in defense of the communist revolution.

Overall, the side storylines provide three key points that complement the main storyline and point towards potential future events. First, the US is likely to be a key player in this narrative, although it is too early at the time of this writing how it might respond. Second, the protests have at least some limited success so far, such as the lifting of customs restrictions. Finally, there is a strong response by the government and its forces, with police and military forces being used, as well as a counter-movement that backs the government.

\textbf{\fontchange{r/Conservative}:}
We present the narrative of \fontchange{r/Conservative}---a right-leaning community centered on US politics \cite{van2020frozen}---in Figure \ref{fig:conservative}. The main story starts with the protests and how Cubans seek to end the regime. However, the rest of the narrative focuses mostly on US political actors. The narrative of \fontchange{r/Conservative} uses the protests as a tool to criticize its political opponents (left-leaning media, Democrats, Black Lives Matter, etc.), rather than focusing on the protests themselves. Regarding the representative landmarks, the first one is a satirical article criticizing communist supporters and their views on the regime after the protests. The second one is part of the main story and highlights how celebrities are also stepping in to support the protests against communism. 


\textbf{\fontchange{r/GenZedong}:}
We present the narrative of \fontchange{r/GenZedong}---a pro-communist community---in Figure \ref{fig:genzedong}. Due to the linear nature of this narrative, extracting representative landmarks is not useful, as they rely on the existence of parallel stories \cite{keith2021narrative}. The storyline starts with an unrelated submission criticizing the presence of an anti-communist Cuban exile in a documentary about Cuba. Although not related to the protests, it aligns with the pro-communist stance of the community. The rest of the timeline supports the regime in Cuba with mostly satirical renditions of anti-communist and pro-US submissions. Overall, this narrative  supports the Cuban regime and criticizes its opponents. 

\textbf{\fontchange{r/worldnews}:}
Figure \ref{fig:worldnews} shows the narrative of \fontchange{r/worldnews}---a neutral community focused on global news \cite{horne2017identifying,bahgat2020towards}. The main storyline starts with an unrelated event on Cuba developing effective vaccines. This is an interesting parallel with the \fontchange{r/cuba} narrative that also started with a health care-related post but in a negative light. The next event is the Mexican president calling for an end to the US embargo---a representative landmark of this map. Next, the storyline focuses on how the protests developed in Cuba. Then, the Miami mayor is discussing bombing Cuba. Immediately after this, we see Cuba lifting the customs restrictions, heavily implying that fear of US intervention led to removing the restrictions, unlike the \fontchange{r/cuba} narrative that presented this as a result of the protests. The side story on the left focuses on the international response (e.g., Russia, China, Iran, and the US). Overall, this narrative has two main themes: the protests themselves and the responses by international actors. This makes sense as the purpose of the community is to discuss world news. Moreover, this external perspective is an interesting complement to the \fontchange{r/cuba} narrative, as it provides a bigger picture view rather than an emotional one. For example, note the lack of rallying call submissions. This makes sense as this community is neutral and should have a more objective view.

\textbf{\fontchange{r/politics}:}
We present the narrative of \fontchange{r/politics}---a left-leaning community centered on US politics \cite{van2020frozen,shepherd2020gaming}--- in Figure \ref{fig:politics}. The narrative starts with a report on the protests, then it mostly focuses on calls to lift the embargo. The storyline then shifts towards the protests in Miami, with a focus on the anti-riot law. The narrative uses this to criticize the political opponents of the community, in a similar way to \fontchange{r/Conservative}.

\begin{table}[!htb]
\resizebox{\columnwidth}{!}{%
\begin{tabular}{@{}p{5.5cm}ccccc@{}}
\toprule
\textbf{Relevant Storylines}             & \textbf{\fontchange{r/cuba}} & \textbf{\fontchange{r/Conservative}} & \textbf{\fontchange{r/GenZedong}} & \textbf{\fontchange{r/worldnews}} & \textbf{\fontchange{r/politics}} \\ \midrule
Causes of the Protest                    & $\times$                     &                                      &                                   & $\times$                          &                                  \\
Cuban Government Response                & $\times$                     &                                      &                                   & $\times$                          & $\times$                         \\
Police Brutality and Repression          & $\times$                     &                                      &                                   & $\times$                          &                                  \\
Supporting the Protests                  & $\times$                     & $\times$                             &                                   &                                   & $\times$                         \\
Supporting the Regime (counter-protests) & $\times$                     &                                      & $\times$                          &                                   &                                  \\
International Response                   &                              &                                      &                                   & $\times$                          &                                  \\
US Response and Potential Intervention   & $\times$                     &                                      & $\times$                          & $\times$                          & $\times$                         \\
Ending the US Trade Embargo              & $\times$                     &                                      & $\times$                          & $\times$                          & $\times$                         \\
Criticizing Political Opponents          & $\times$                     & $\times$                             & $\times$                          &                                   & $\times$                         \\ \bottomrule
\end{tabular}%
}
\caption{Relevant storylines \& events in each narrative map.} 
\label{tab:narratives}
\vspace{-25pt}
\end{table}

\textbf{\fontchange{Narrative Comparison}:} We show the relevant storylines and events in each narrative in Table \ref{tab:narratives}. We note that \fontchange{r/cuba} covers almost all the aspects of the protest except the international response, save for the US, given its relevance in the historical and geopolitical context of Cuba. In contrast, the other communities tend to focus on specific storylines based on their agenda. 


\section{Discussion}
\textbf{Social Movement Narratives.}
Our work unravels the different narratives espoused by online communities pertaining to the 2021 Cuban protest movement. In general, our work provides a computational framework to analyze the accepted narratives that emerge in online communities in the context of social movements, allowing analysts and social scientists to understand the motivations and relations between different groups in the context of a social movement. In particular, we found that \fontchange{r/cuba} had the most complex narrative with a tree-like structure and diverse content covering multiple storylines. In contrast, \fontchange{r/GenZedong} had the simplest narrative---a single linear storyline---and had pro-communist content. Meanwhile, both the US-based political communities---\fontchange{r/politics} and \fontchange{r/Conservative}---used the protests to criticize their opponents and further their own agendas, rather than focusing on the protests themselves. Finally, \fontchange{r/worldnews} showed the most neutral presentation, focusing on factual content rather than emotional content, with news discussing the international response and the events of Cuba itself.

\textbf{Implications.}
In the context of computational journalism \cite{cohen2011computational}---which provides tools for various computational needs in journalism---our approach could serve as a support tool for journalists. For example, narrative maps could be part of a narrative exploration suite, which could also include text summaries, data plots, and references to trustworthy sources. Such a tool could aid analysts, journalists, and other users in trying to understand the narratives espoused by online communities \cite{keith2021narrative}. Our work could also be applied by social scientists interested in a narrative analysis of how social movements are evolving, as the accepted narratives provide a window into what different communities are thinking and their views. 

Furthermore, our community acceptance approach could be enhanced with additional attributes. In particular, we could model the credibility and political bias of Reddit submissions (e.g., by analyzing the source they are linking). These attributes could aid in modeling how misinformation spreads during social movements. Another possible extension would be using sentiment analysis \cite{ravi2015survey} in conjunction with our narrative model to obtain more nuanced narratives that explicitly model how communities feel with respect to specific storylines and events. Finally, we note that our methods are grounded in theoretical constructs of narratology \cite{keith2021narrative} and could be applied beyond social movement narratives.

\textbf{Ethical Considerations.} 
There are two stakeholder groups in this research: members of the studied communities and potential users of the system (e.g., analysts, journalists, and researchers). As with any system that could be used to monitor social movements, there is a risk that malicious actors leverage the system for nefarious uses, such as tracking activists involved in organizing or participating in these movements or designing targeted information operations to counter or support specific narratives. Regarding tracking issues, the system works at an aggregate level, without user-level details, and the system uses non-confidential and publicly available data and does not require any specific user information. Thus, it would be hard to perform user re-identification. Next, we note that the system does not distinguish between the truth and misinformation, which could allow users to explore how misinformation spreads in online communities, but also lead to incorrect interpretations of the narratives. We could reduce the amount of potential misinformation in maps by giving control to the analysts over the information sources used. 

Nevertheless, we note that the narrative extraction tool is still in the prototyping phase. We expect to address any issues that arise as we work with stakeholders in user studies and pilot testing. Through this process, we will ensure that the system is resistant to these issues. Finally, we note that our approach does not involve any direct interaction with the studied communities, it only analyzes publicly available content from these communities. Thus, we do not anticipate any direct harm resulting from this research.

\textbf{Limitations.} Our work is not without limitations. The current embedding model only considers the text of a submission, disregarding images and videos. Thus, future work could consider multimedia embeddings to model these elements. Moreover, we note that our data spans a relatively small period of time. Thus, the analysis does not capture all the narratives surrounding the movement.

\section{Conclusions}
This paper presents an algorithmic approach to extract the accepted narratives of online communities in the context of social movements, specifically the 2021 Cuban Protests during its first few days. Building upon existing narrative extraction methods based on linear optimization, we designed a novel approach that integrates the concept of \textit{community acceptance} with pre-existing definitions of coherence and coverage. Our technique allows us to identify the major themes surrounding the protests in five different online communities. Recognizing the accepted narratives of different communities is central to understanding social movements. Thus, our method could aid journalists in making sense of emerging narratives in social media in the context of reporting social movements. In general, our work has potential applications in the study of social movements, intelligence analysis, computational journalism, and misinformation research. Moreover, it opens up multiple avenues for future research, such as adapting the method to find rejected narratives or modeling acceptance in other social networks.

\section*{Acknowledgments}
This work was supported in part by NSF I/UCRC CNS-1822080 via the NSF Center for Space, High-performance, and Resilient Computing (SHREC), the NSF grants CNS-1915755 and DMS-1830501, and by ANID/Doctorado Becas Chile/2019 - 72200105.

\bibliographystyle{ACM-Reference-Format}
\bibliography{acl2019}

\newpage
\appendix
\onecolumn
\section*{Appendix A: Linear Model Details}

\begin{table}[H]
\centering
\resizebox{\textwidth}{!}{%
\begin{tabular}{@{}p{2.2cm}lp{13cm}@{}}
\toprule
\textbf{Type}                    & \textbf{Name}                           & \textbf{Description}                                                                           \\ \midrule
\textbf{Variables}               & $\mathsf{minedge}$                      & Represents the lowest coherence value over all active edges.                                       \\
                                 & $\mathsf{event}_{i}$                    & Represents the presence of event $i$ in the narrative map.                                         \\
                                 & $\mathsf{edge}_{ij}$                    & Represents the strength of the connection between events $i$ and $j$ in the narrative map.         \\
                                 & $\mathsf{cluster}_{k}$                  & Represents the average presence of cluster $k$ in the narrative map.                              \\ \midrule
\multirow[t]{6}{2.2cm}{\textbf{Data-based Parameters}}   & $\mathsf{num\mhyphen clusters}$         & Represents the number of clusters of events.                                                 \\
                                 & $\mathsf{membership}_{ijk}$             & Represents whether the connection of events $i$ and $j$ belongs to cluster $k$. Defined as $\sqrt{\mathsf{membership}_{ik} \cdot \mathsf{membership}_{jk}}$---the geometric mean between the probabilities of the events belonging to cluster $k$.                                         \\
                                 & $\mathsf{coherence}_{ij}$               & Represents the coherence score of the connection between events $i$ and $j$. Defined as $\sqrt{\mathsf{angular\mhyphen similarity}_{ij} \cdot \mathsf{cluster\mhyphen similarity}}_{ij}$---the geometric mean between the angular similarity of the underlying event embeddings and the clustering similarity based on comparing the cluster membership probability distributions of each event using Jensen-Shannon divergence.               \\
                                 & $\mathsf{score\mhyphen percentile}_{i}$ (*) & Represents the score percentile of event (submission) $i$ in the community.                                               \\
                                 & $\mathsf{upvote\mhyphen ratio}_i$ (*)& Represents the upvote ratio of event (submission) $i$ in the community.                                                   \\
                                 & $\mathsf{acceptance}_{ij}$ (*)             & Represents the acceptance score of the connection between events $i$ and $j$. Defined as $\sqrt{\mathsf{score\mhyphen percentile}_{i} \cdot \mathsf{score\mhyphen percentile}_{j}} \cdot \sqrt{\mathsf{upvote\mhyphen ratio}_i \cdot \mathsf{upvote\mhyphen ratio}_j}$---the product between the geometric means of submission score percentiles and upvote ratios.                                 \\ \midrule
\multirow[t]{3}{2.2cm}{\textbf{User-defined Parameters}} & $K$                                     & Defines the expected length of the main storyline in the narrative map.                         \\
                                 & $\mathsf{mincover}$                     & Defines the minimum average coverage over all clusters in the narrative map.                       \\
                                 & $\mathsf{minscore}$ (*)                     & Defines the minimum average acceptance over all events (submissions) in the narrative map.                                \\ \bottomrule
\end{tabular}%
}
\caption{Description of the variables and parameters of the proposed extraction algorithm. All elements take values between $0$ and $1$, except $K$ and $\mathsf{num\mhyphen clusters}$, which are positive integers. Most elements of the model are based on the extraction algorithm of \citet*{keith2021narrative}. The (*) symbol represents new additions to the linear problem formulation designed to incorporate \textit{community acceptance} in social media into the model.}
\label{tab:lp_model}
\end{table}

\newpage
\twocolumn
\section*{Appendix B: Narrative Map Figures}

\begin{figure}[H]
    \centering
    \includegraphics[width=0.85\columnwidth]{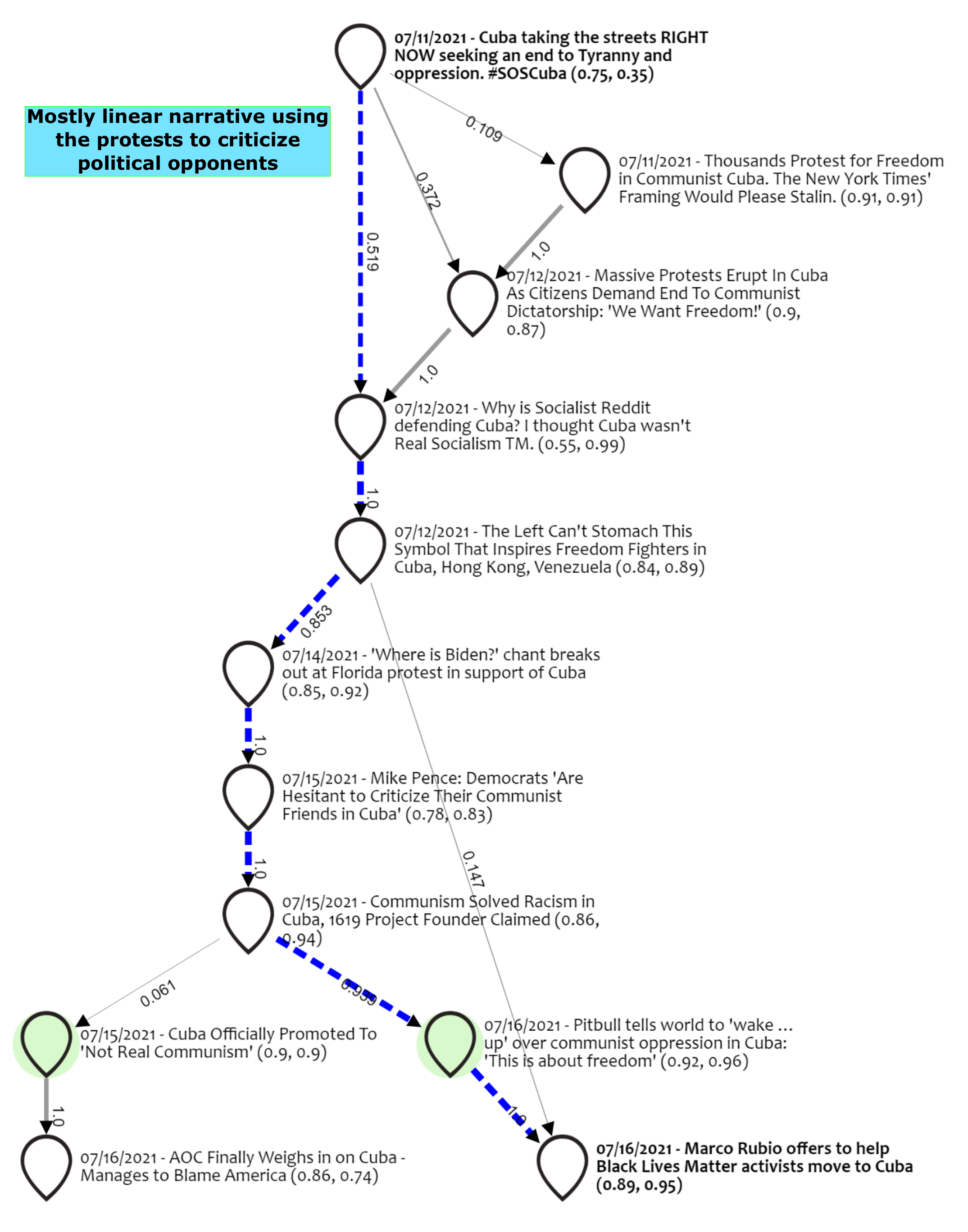}
    \caption{Narrative map for \fontchange{r/Conservative}.}
    \label{fig:conservative}
\end{figure}

\begin{figure}[H]
    \centering
    \includegraphics[width=0.55\columnwidth]{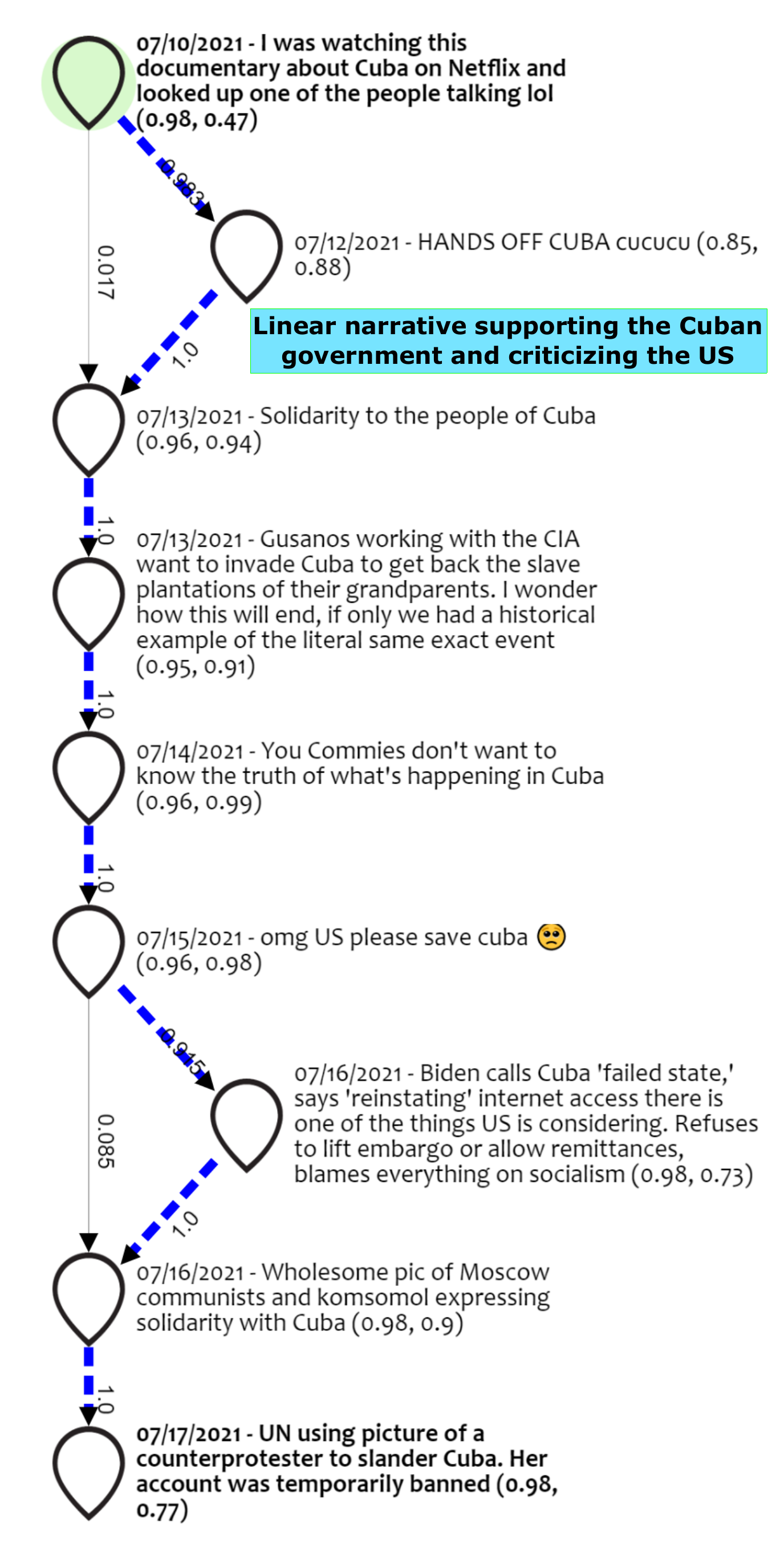}
    \caption{Narrative map for \fontchange{r/GenZedong}.}
    \label{fig:genzedong}
\end{figure}

\begin{figure}[H]
    \centering
    \includegraphics[width=0.85\columnwidth]{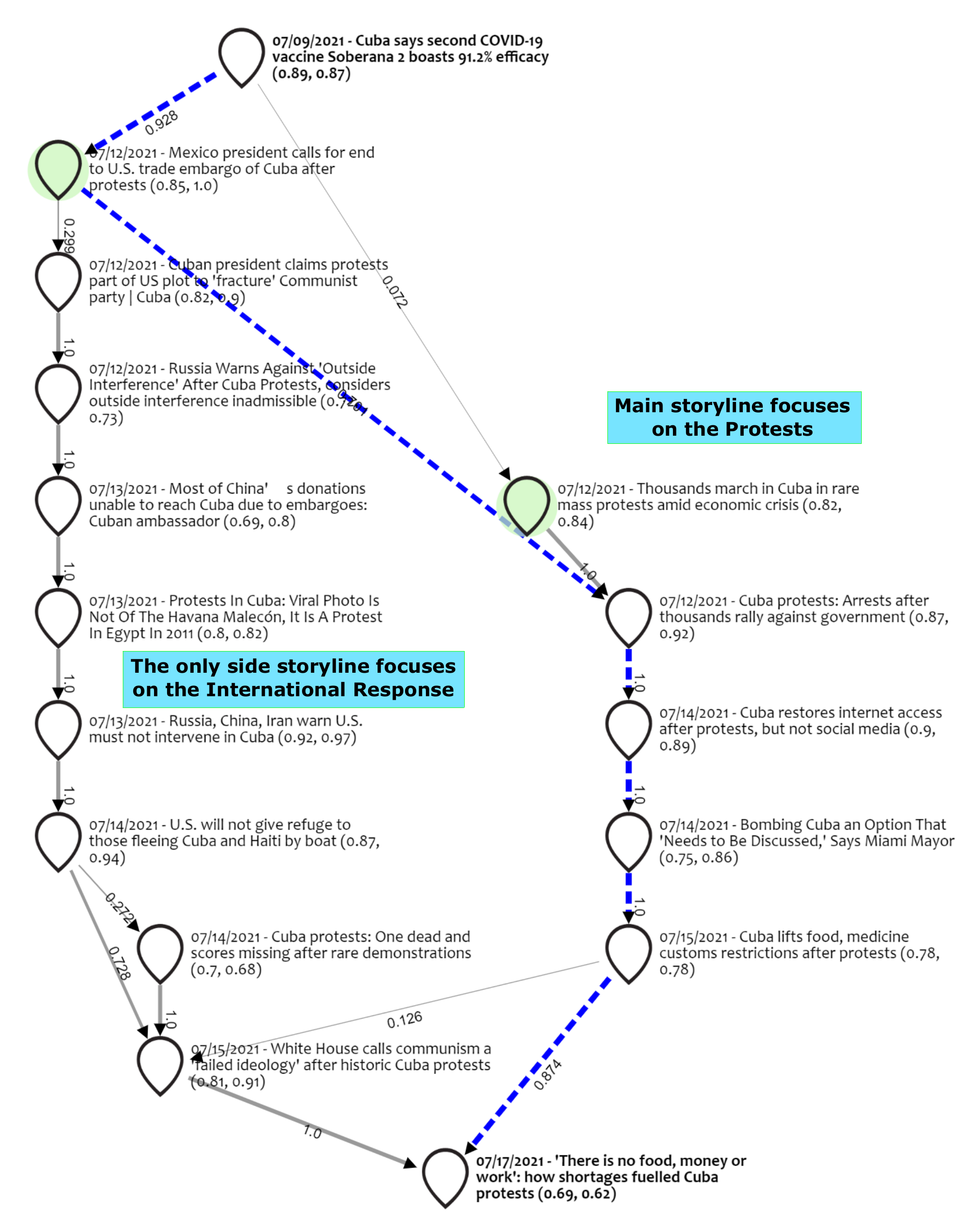}
    \caption{Narrative map for \fontchange{r/worldnews}.}
    \label{fig:worldnews}
\end{figure}

\begin{figure}[H]
    \centering
    \includegraphics[width=\columnwidth]{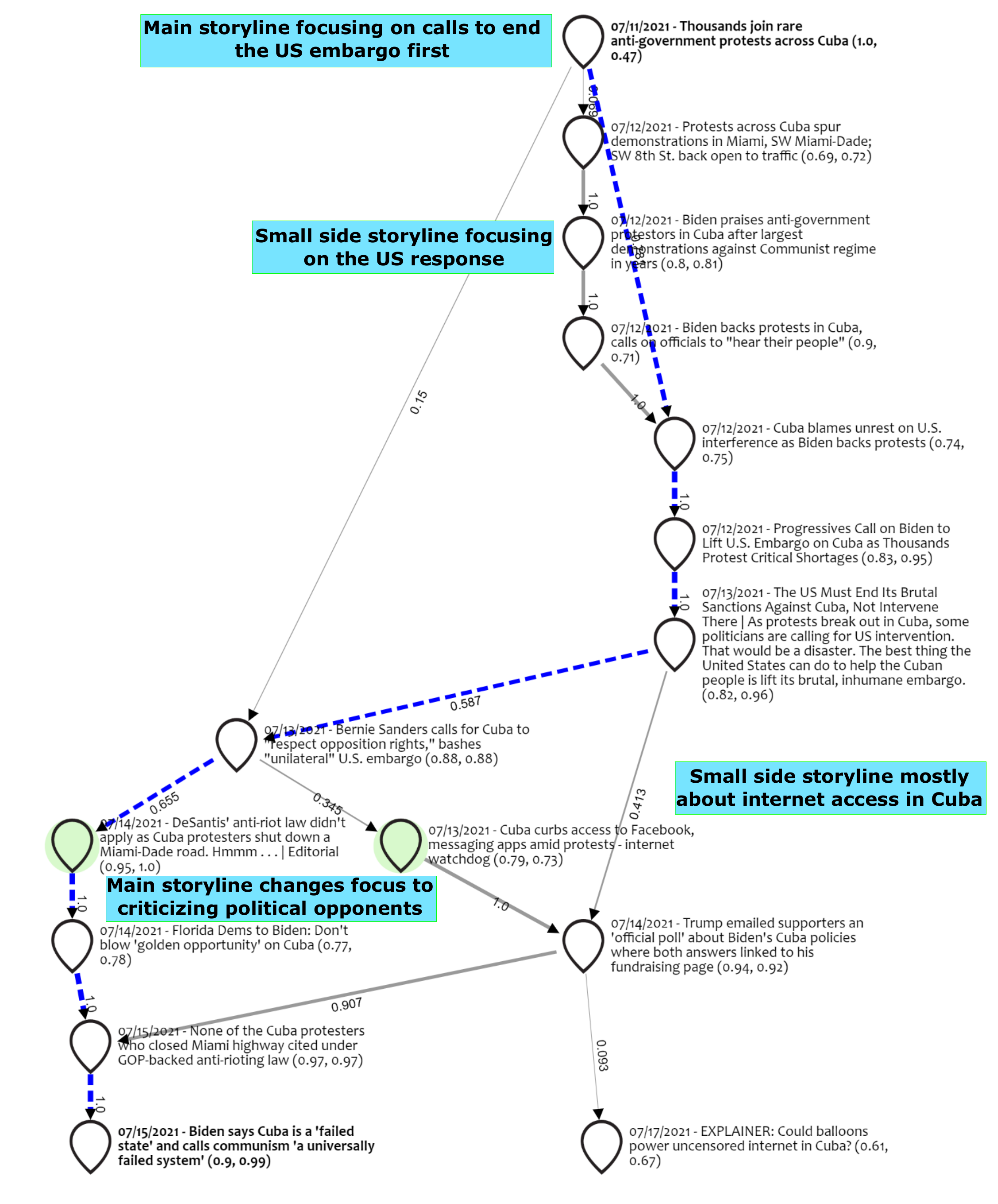}
    \caption{Narrative map for \fontchange{r/politics}.}
    \label{fig:politics}
\end{figure}

\end{document}